\documentclass[12pt,a4paper]{article}
\usepackage{a4wide}
\usepackage{latexsym}
\usepackage{epsf}
\usepackage{amssymb}
\usepackage{graphicx}
\usepackage{amsmath, cite}
\usepackage{slashed,epsfig}
\include{bracket}
\usepackage{amsfonts}
\usepackage{bbm}
\usepackage{bbm}
\usepackage{amsmath,amssymb,amsthm}
\usepackage{pst-all}
\usepackage{here}
\usepackage{latexsym,amssymb}
\usepackage{mcite}

 \makeatletter
\@addtoreset{equation}{section} \makeatother

\def\bfone{\relax{\rm 1\kern-.35em 1}}

 \makeatletter
\@addtoreset{equation}{section} \makeatother

\def\be {\begin{equation}}
\def\ee {\end{equation}}
\def\bea {\begin{eqnarray}}
\def\eea {\end{eqnarray}}
\def\bc {\begin{center}}
\def\ec {\end{center}}
\def\a  {\alpha}

\def\D  {\Delta}

\def\bfg {\begin{figure}}
\def\efg {\end{figure}}
\def\bi {\begin{itemize}}
\def\ei {\end{itemize}}
\def\nn {\nonumber}

\def\la {\label}
\def\le {\left}
\def\ri {\right}

\DeclareFontFamily{U}{rsf}{} \DeclareFontShape{U}{rsf}{m}{n}{
  <5> <6> rsfs5 <7> <8> <9> rsfs7 <10-> rsfs10}{}
\DeclareMathAlphabet\Scr{U}{rsf}{m}{n}

\begin{document}

\begin{center}
{\bf \large{ Minimal Length in Quantum Gravity, Equivalence Principle and Holographic Entropy Bound\\[10mm]}}
\large {Ahmed Farag Ali} \\[4mm]
\small{Dept. of Physics, University of Lethbridge, Lethbridge, Alberta, Canada T1K 3M4}\\[4mm]
{\upshape{ahmed.ali@uleth.ca}}\\[7mm]
\end{center}

\begin{abstract}
A possible discrepancy has been found between the results of a neutron interferometry
experiment and Quantum Mechanics. This experiment suggests that the weak
equivalence principle is violated at small length scales, which quantum mechanics
cannot explain. In this paper, we investigated whether the Generalized Uncertainty
Principle (GUP), proposed by some approaches to quantum gravity such as String
Theory and Doubly Special Relativity Theories (DSR), can explain the violation
of the weak equivalence principle at small length scales. We also investigated the
consequences of the GUP on the Liouville theorem in statistical mechanics.
We have found a new form of invariant phase space in the presence of GUP.
This result should modify the density states and affect the calculation
of the entropy bound of local quantum field theory, the cosmological constant,
black body radiation, etc. Furthermore, such modification may have observable
consequences at length scales much larger than the Planck scale. This modification
leads to a $\sqrt{A}$-type correction  to  the bound of the maximal entropy
of a bosonic field which would definitely shed some light on the holographic theory.
\end{abstract}


\section{Inroduction}

The existence of a minimal length is one of the most interesting predictions
of some approaches related to quantum gravity such as String Theory
and Black hole physics. This is a consequence of Perturbation String
Theory since strings can not interact at distances smaller than their size.
One of the interesting phenomenological
implications of the existence of the minimal measurable length is
the modification of the standard commutation relation, between
position and momentum, in usual quantum mechanics to the so-called
generalized uncertainty principle (GUP). Recently,
we proposed the GUP  in \cite{advplb,Das:2010zf} which
is consistent with  Doubly Special Relativity (DSR) theories,
String Theory and Black Holes Physics {\it and}
which ensures $[x_i,x_j]=0=[p_i,p_j]$ (via the Jacobi identity).

\bea
[x_i, p_j]\hspace{-1ex} &=&\hspace{-1ex} i \hbar\hspace{-0.5ex} \left[  \delta_{ij}\hspace{-0.5ex}
- \hspace{-0.5ex} \alpha \hspace{-0.5ex}  \le( p \delta_{ij} +
\frac{p_i p_j}{p} \ri)
+ \alpha^2 \hspace{-0.5ex}
\le( p^2 \delta_{ij}  + 3 p_{i} p_{j} \ri) \hspace{-0.5ex} \ri]
\label{comm01} \\
%
%
%
 \Delta x \D p \hspace{-1ex}&\geq &\hspace{-1ex}\frac{\hbar}{2}
\le[ 1 - 2 \alpha <p> + 4\alpha^2 <p^2>
\ri] ~ \nn \\
\hspace{-1ex}&\geq& \hspace{-1ex}
\frac{\hbar}{2} \hspace{-1ex}
\le[\hspace{-0.5ex} 1\hspace{-0.5ex}  +\hspace{-0.7ex}  \le(\hspace{-0.7ex}  \frac{\alpha}{\sqrt{\langle p^2 \rangle}} +\hspace{-0.2ex}4\alpha^2 \hspace{-0.9ex} \ri)
\hspace{-0.6ex}  \D p^2 \hspace{-0.6ex}
+\hspace{-0.6ex}  4\alpha^2 \langle p \rangle^2 \hspace{-0.6ex}
- \hspace{-0.6ex}  2\alpha \sqrt{\langle p^2 \rangle}\hspace{-0.2ex}
\ri]\hspace{2ex} \label{dxdp1}
\eea
where
$\alpha = {\alpha_0}/{M_{Pl}c} = {\alpha_0 \ell_{Pl}}/{\hbar},$
$M_{Pl}=$ Planck mass, $\ell_{Pl}\approx 10^{-35}~m=$ Planck length,
and $M_{Pl} c^2=$ Planck energy $\approx 10^{19}~GeV$.
Various versions of the GUP have been proposed by many authors,
motivated by String Theory, Black Hole Physics, DSR etc, see e.g.
\cite{guppapers,kmm,kempf,brau,sm,cg}, and for investigating
phenomenological implications see \cite{dvprl,dvcjp,Ali:2010yn}.) Note
that  Eqs. (\ref{comm01}) and (\ref{dxdp1}) are approximately covariant under
DSR transformations \cite{cg}. Since DSR transformations
preserve both speed of light, and  invariant energy scale,
it is not surprising that Eqs. (\ref{comm01}) and (\ref{dxdp1})
imply the existence of minimum measurable length and
maximum measurable momentum
\bea
\D x &\geq& (\D x)_{min}  \approx \alpha_0\ell_{Pl} \la{dxmin} \\
\D p &\leq& (\D p)_{max} \approx \frac{M_{Pl}c}{\alpha_0}~. \la{dpmax}
\eea
\par\noindent
It can be shown that the following definitions
\bea x_i = x_{0i}~,~~
p_i = p_{0i} \le( 1 - \alpha p_0 + 2\alpha^2 p_0^2 \ri)~, \la{mom1}
\eea
(with $x_{0i}, p_{0j}$
satisfying the canonical commutation relations $ [x_{0i}, p_{0j}] = i \hbar~\delta_{ij}, $
such that $p_{0i} = -i\hbar \partial/\partial{x_{0i}}$) satisfy Eq.(\ref{comm01}).
In \cite{advplb}, we have shown that any non-relativistic  Hamiltonian of the form
$H=p^2/2m + V(\vec r)$ can be written as
$H = p_0^2/2m - (\alpha/m)p_0^3 + + V(r) + {\cal O}(\alpha^2)$ using
Eq.(\ref{mom1}), where the second term can be treated as a perturbation. Now, the
third order Schr\"odinger equation has a new {\it non-perturbative}
solution of the form $\psi \sim e^{ix/2a\hbar}$. When
applied to an elementary particle, it implies that the space
 which confines it must be discrete.

\bea
\frac{L}{a\hbar} = \frac{L}{a_0 \ell_{Pl}} = 2p\pi + \theta~,~p \in \mathbb{N}
\la{quant1}
\eea

This suggests that space itself is discrete, and that all measurable lengths are quantized
in units of a fundamental minimum measurable length (which can be the Planck length).

Considering the relativistic case in \cite{Das:2010zf} was important for many reasons.
The relativistic particles are natural candidates for studying the nature of
spacetime near the Planck scale. Also, most of the elementary particles in the nature
are fermions, obeying some form of the Dirac equation. Furthermore, It is easier
to investigate whether the discreteness of space exist in $2$ and $3$ dimensions
by studying Dirac equation with GUP. We have shown that to confine the particle in the
D-dimensional box, the dimensions of the box  would have to be quantized in multiples
of a fundamental length, which can be  the Planck length.

The scope of the present work is to investigate the effect of Quantum Gravity
Corrections on the equivalence principle and the holographic entropy bound.
In section $2$, we investigate Heisenberg equations of motion
in the presence of the GUP, we found that the acceleration is no longer
mass-independent because of the mass-dependence through the momentum p.
Therefore, the equivalence principle is dynamically violated.
In section $3$, we tackle a naturally arising question of whether
the number of states inside a volume of phase space does not change with
time in the presence of the GUP. So, we calculate the consequences
of the GUP on the Liouville theorem in statistical mechanics. We applied our approach
on the entropy bound of local quantum field theory. This leads
to a $\sqrt{A}$-type correction  to the bound of the
maximal entropy of a quantum field.


\section{The Equivalence principle at short distance}

Quantum Mechanics does not violate Equivalence  principle. This can be shown
from studying Heisenberg equations of motion.
For simplicity, consider $1-dimensional$ motion with the Hamiltonian given by
\be
H= \frac{P^2}{2m}+V(x).
\ee

The Heisenberg equations of motion read,
\bea
\dot{x}&=&\frac{1}{i \hbar}[x,H]= \frac{p}{m},\\
\dot{p}&=&\frac{1}{i \hbar}[p,H]=-\frac{\partial V}{\partial x}.
\eea

These equations ensure that the momentum at the quantum level is $p=m\dot{x}$
and the acceleration $\ddot x$ is  mass-independent like in classical physics.
It is obvious that the equivalence principle is preserved at the quantum level,
and it is clear that this result possibly contradicts experimental results \cite{exp}.

Let us study Eq(\ref{comm01}) at the classical limit using the correspondence between
commutator in quantum mechanics and poisson bracket in classical mechanics,
\begin{equation}
\frac{1}{i\hbar} [ \hat{P}, \hat{Q} ] \Longrightarrow \{P,Q\}\;,
\end{equation}
so the classical limit of Eq(\ref{comm01}) give

\begin{eqnarray}
\{x_i,p_j\} & = &  \delta_{ij} - \a( p \delta_{ij} + \frac{p_i p_j}{p}) + \a^2 (p^2 \delta_{ij}+ 3 p_i p_j).
\label{Eq:Poi1}
\end{eqnarray}

The equations of motion are given by

\begin{eqnarray}
\dot{x}_i & = & \{x_i,H\}
\;=\; \phantom{-}\{x_i,p_j\}\,\frac{\partial H}{\partial p_j}
     \;,\cr
\dot{p}_i & = & \{p_i,H\}
\;=\; -\{x_j,p_i\}\,\frac{\partial H}{\partial x_j} \;.
\end{eqnarray}

Consider the effect of the GUP on $1-dimensional$ motion with the Hamiltonian given by,
\be
H= \frac{P^2}{2m}+V(x).
\ee
The  equations of motion will be modified as follows,
\bea
\dot{x}&=&\{x,H\}= (1-2\a p)\frac{p}{m} \label{eq1},\\
\dot{p}&=&\{p,H\}=(1-2\a p)(-\frac{\partial V}{\partial x}),\label{eq2}
\eea
where the momentum p is no longer equal to $m \dot{x}$.

Using (\ref{eq1},\ref{eq2}), we can derive the acceleration given by,
\be
\ddot{x}=-(1-6\a p) \frac{\partial V}{\partial x}.
\ee

Notice that if the force $F=  - \frac{\partial V}{\partial x}$ is gravitational
and proportional to the mass $m$, the acceleration $\ddot x$ is not
mass-independent because of the  mass-dependence through the momentum $p$.
Therefore, the equivalence principle is dynamically violated because of
generalized uncertainty principle. Since the GUP is an aspect of various
approaches to Quantum Gravity such as  String Theory and Doubly
Special Relativity (or DSR) Theories, as well as black hole physics,
it is promising to predict the upper bounds on the quantum gravity
parameter compatible with the experiment that was done in \cite{exp}.
This result agrees, too, with cosmological implications of the dark sector where
a long-range force acting only between nonbaryonic particles would be associated with a large
violation of the weak equivalence principle \cite{exp2}.
The violation of equivalence principle has been obtained, too, in the context
of string theory\cite{ST} where the extended nature of strings
are subject to tidal forces and do not follow geodesics.

\section{ The GUP and Liouville theorem }

In this section, we continue our investigation of the consequences
of  our proposed commutation relation of Eq(\ref{comm01}).
What we are looking for is an analog of the Liouville theorem
in presence of the GUP. We should make sure that the number of states inside
volume of phase space does not change with time revolution in presence of the GUP. If this
is the case, this should modify the density states and affect the
Entropy bound of local quantum field theory, the Cosmological constant,
black body radiation, etc. Furthermore, such modification  may have
observable consequences at length scales much larger than the Planck scale.
The Liouville theorem has been studied before with different versions of GUP, see e.g.
\cite{Chang:2001bm}.

Since we are seeking the number of states inside a volume of phase space
that does not change with time, we assume the time evolutions
of the position and momentum during $\delta t$ are

\begin{eqnarray}
x_i' & = & x_i + \delta x_i \;, \cr
p_i' & = & p_i + \delta p_i \;,\la{time}
\end{eqnarray}

where

\begin{eqnarray}
\delta x_i
& = &  \{x_i,p_j\}\,\frac{\partial H}{\partial p_j}
           \delta t \;, \cr
\delta p_i
& = &-\{x_j,p_i\}\,\frac{\partial H}{\partial x_j}
      \delta t \;.\la{delta}
\end{eqnarray}

The infinitesimal phase space volume after this infinitesimal time evolution is

\begin{equation}
d^D\mathbf{x}'\,d^D\mathbf{p}'
= \left| \dfrac{\partial(x'_1,\cdots,x'_D,p'_1,\cdots,p'_D)}
               {\partial(x_1, \cdots,x_D, p_1, \cdots,p_D)}
  \right|
d^D\mathbf{x}\,d^D\mathbf{p} \;.
\end{equation}

Using Eq(\ref{time}),  the Jacobian reads  to the  first order in $\delta t$

\begin{equation}
\left| \dfrac{\partial(x'_1,\cdots,x'_D,p'_1,\cdots,p'_D)}
               {\partial(x_1, \cdots,x_D, p_1, \cdots,p_D)}
\right| =
1 + \left( \frac{\partial\delta x_i}{\partial x_i}
         + \frac{\partial\delta p_i}{\partial p_i}
    \right)
  + \cdots \;.
\end{equation}
Using Eq(\ref{delta}), we get
\begin{eqnarray}
\lefteqn{
\left(\frac{\partial\delta x_i}{\partial x_i}
    + \frac{\partial\delta p_i}{\partial p_i}
\right)\frac{1}{\delta t} } \cr
& = & \frac{\partial}{\partial x_i}
      \left[ \{x_i,p_j\}\,\frac{\partial H}{\partial p_j}
           \right]
    - \frac{\partial}{\partial p_i}
      \left[ \{x_j,p_i\}\,\frac{\partial H}{\partial x_j}
      \right] \cr
& = & \left[ \frac{\partial}{\partial x_i}\{x_i,p_j\} \right]
      \frac{\partial H}{\partial p_j}
    + \{x_i,p_j\}\frac{\partial^2 H}{\partial x_i \partial p_j}
    \frac{\partial H}{\partial x_j} \cr
& & - \left[ \frac{\partial}{\partial p_i}\{x_j,p_i\} \right]
      \frac{\partial H}{\partial x_j}
    - \{x_j,p_i\}\frac{\partial^2 H}{\partial p_j \partial x_i} \cr
& = & - \left[ \frac{\partial}{\partial p_i}\{x_j,p_i\} \right]
      \frac{\partial H}{\partial x_j} \cr
& = &- \frac{\partial}{\partial p_i}\left[\delta_{ij}- \a (p \delta_{ij}+\frac{p_i p_j}{p})\right] \frac{\partial H}{\partial x_j} \cr
& = & \frac{\partial}{\partial p_i} \a (p \delta_{ij}+\frac{p_i p_j}{p}) \frac{\partial H}{\partial x_j} \cr
& = & \a (D+1) \frac{p_j}{p} \frac{\partial H}{\partial x_j}.
\end{eqnarray}

The  infinitesimal phase space volume after this infinitesimal evolution up to first order in $\a $ and $\delta t$ is

\begin{equation}
d^D\mathbf{x}'\,d^D\mathbf{p}'
= d^D\mathbf{x}\,d^D\mathbf{p}\left[ 1+ \a (D+1)  \frac{p_i}{p} \frac{\partial H}{\partial x_i}  \delta t \right].
\end{equation}

Now we are seeking  the analog of the Liouville theorem in which the weighted phase space volume
is invariant under time evolution.
Let us check the infinitesimal evolution  of $(1-\a p^{\prime})$ up to first order in $\a$ and $\delta t$

\begin{eqnarray}
(1 - \a p^{\prime}) &=& 1-\a\sqrt{p_i^\prime p_i^\prime}\cr
&=& 1-\a \Big[(p_i+\delta p_i) (p_i+\delta p_i)\Big]^{\frac{1}{2}}\cr
&\approx& 1-\a (p^2+2 p_i\delta p_i)^{\frac{1}{2}}\cr
&\approx& 1-\a\Big[p^2- 2 p_i \{x_i,p_j\}\frac{\partial H}{\partial x_j}\delta t  \Big]^\frac{1}{2}\cr
&\approx& 1- \a \Big[p-\frac{1}{p}(p_j-2 \a p p_j)\frac{\partial H}{\partial x_j} \delta t \Big]\cr
&\approx& (1- \a p)+ \a \frac{p_j}{p}(1-2 \a p) \frac{\partial H}{\partial x_j} \delta t\cr
&\approx& (1-\a p)\Big[1 +\a \frac{p_j}{p} \frac{1-2 \a p}{1-\a p} \frac{\partial H}{\partial x_j} \delta t\Big]\cr
&\approx& (1-\a p) \Big[1+  \a \frac{p_j}{p} \frac{\partial H}{\partial x_j} \delta t\Big].
\end{eqnarray}

Therefore, we get to first order in $\a$ and $\delta t$,
\begin{equation}
(1- \a p^\prime)^{-D-1}= (1-\a p)^{-D-1}\Big[1- (D+1) \a \frac{p_j}{p}\frac{\partial H}{\partial x_j} \delta t \Big]
\end{equation}

This result in the following expression is invariant under time evolution!

\begin{equation}
\frac{ d^D\mathbf{x^\prime}\,d^D\mathbf{p^\prime} }
     { (1-\a p^\prime)^{D+1} }=\frac{ d^D\mathbf{x}\,d^D\mathbf{p} }
     { (1-\a p)^{D+1} } \,.
\label{INVARIANT}
\end{equation}

If we integrate over the coordinates, the invariant phase space volume  of Eq.~(\ref{INVARIANT}) will be

\begin{equation}
\frac{V\,d^D\mathbf{p}}{(1- \a p)^{D+1}} \;,
\end{equation}
 Where $V$ is the coordinate space volume.
 The number of quantum states per momentum space volume can  be assumed to be

\begin{equation}
\frac{V}{(2\pi\hbar)^D}\frac{d^D\mathbf{p}}{(1-\a p)^{D+1}} \;.
\label{Densitystates}
\end{equation}

The modification  in the number of quantum states per momentum space volume
in (\ref{Densitystates}) should have consequences on  the calculation of the
entropy bound of local quantum field theory, the cosmological constant, black body
radiation, etc. In this paper, we are investigating its consequences on the
entropy bound of local quantum field. In the following two subsections we  briefly
introduce the  holographic entropy  bound proposed by't Hooft \cite{thooft} and
entropy bound of Local quantum field proposed by Yurtsever and Aste
\cite{Yurt,Aste06,Aste04}. In subsection(3.2)  we treat the
effects of the GUP  on the entropy bound of Local quantum field.

\subsection{The Holographic Entropy Bound and Local Quantum Field Theory }
The entropy of a closed spacelike surface containing quantum  bosonic field
has been studied by 't Hooft \cite{thooft} . For the field states to be observable
for outside world, 't Hooft assumed that their energy inside the surface
should be less than $1/4$  times its linear dimensions, otherwise the
surface would lie within the Schwarzschild radius\cite{thooft}.\\

If the bosonic quantum fields are confined to closed spacelike surface at a temperature T,
the energy of the most probable state is

\be
E= a_1 Z T^4 V, \label{energyt}
\ee
where $Z$ is the number of different fundamental particle types with mass less than T and
$a_1$ a numerical constant of order one, all in natural units.\\

Now turning to the total entropy $S$, it is found that it is given by

\be
S=a_2 Z V T^3,
\ee
where $a_2$ is another numeric constant of order one.\\

The Schwarzschild limit requires that
\be
2 E <\frac{V}{\frac{4}{3} \pi}.
\ee

Using Eq(\ref{energyt}), one finds

\be
T<a_3 Z^{-\frac{1}{4}} V^{-\frac{1}{6}},
\ee

so the entropy bound is given by
\be
S<a_4  Z^{\frac{1}{4}} V^{\frac{1}{2}}= a_4  Z^{\frac{1}{4}} A^{\frac{3}{4}},
\ee
where $A $ is the boundary area of the system.
At low temperatures, Z is limited by a dimensionless number
, then this entropy is small compared to that
of a black hole, if the area A is sufficiently large.
The black hole is the limit of maximum entropy
\be
S_{max}=\frac{1}{4} A.
\ee

Therefore, for any closed surface without worrying
about its geometry inside, all physics can be represented by degrees
of freedom on this surface itself. This implies that the quantum gravity can be described
by a topological quantum field theory, for which all physical degrees of freedom
can be projected onto the boundary\cite{thooft}. This is know as Holographic Principle.\\

According to Yurtsever's  paper\cite{Yurt}, the holographic
entropy bound can be derived from elementary flat-spacetime quantum field theory when
the total energy of Fock states is in a stable configuration against
gravitational collapse by imposing a cutoff on the maximum energy of the field
modes of the order of the Planck energy. This leads to an entropy bound of holographic type.

Consider a massless bosonic field confined to  cubic
box of size $L$ , as has been done  in~\cite{Kim:2008kc,Yurt,Aste04,Aste06,CX},
the total number of the quantized modes is given by
\begin{equation}
 N=\sum_{\vec{k}}1 \rightarrow \frac{L^3}{(2\pi)^3}\int d^3\vec{p}
  =\frac{L^3}{2\pi^2}\int_{0}^{\Lambda}p^2 dp
  =\frac{\Lambda^3L^3 }{6\pi^2},
\end{equation}
where $\Lambda$ is the UV energy cutoff of the
LQFT.   The UV cutoff makes $N$  finite. The Fock states can be
constructed by assigning occupying number $n_i$ to these $N$ different modes
\begin{equation}
 \mid\Psi\!>=\mid n(\vec{k}_1), n(\vec{k}_2),\cdots,n(\vec{k}_N)>
 ~~\to~~ \mid n_1,n_2,\cdots,n_N>,
\end{equation}
The dimension of the Hilbert space is calculated by the  number of occupancies
 $\{n_i\}$ which is finite if it is bounded. The non-gravitational
collapse condition leads to finiteness of the Hilbert space.
\begin{equation} \label{energy}
E= \sum_{i=1}^N n_i \omega_i  \leq
 E_{BH}=L.
\end{equation}

It can be observed that $N$ particle state with one particle occupying
one mode ($n_i=1$) corresponds to the lowest energy state with $N$
modes simultaneously excited. In this case, it should satisfy the
gravitational stability condition of Eq.(\ref{energy}).
Hence, the energy bound is given by

\begin{equation}
E \rightarrow
\frac{L^3}{2\pi^2}\int_{0}^{\Lambda} p^3 dp =\frac{\Lambda^4 L^3}{8\pi^2} \leq
 E_{BH}.
\end{equation}

The last inequality implies

\begin{equation} \label{UVR}
\Lambda^2\leq \frac{1}{L}.
\end{equation}

The maximum entropy is given by
\begin{equation}
S_{\rm max}=-\sum_{j=1}^W \frac{1}{W}\ln\frac{1}{W}= \ln W\label{maxentropy},
\end{equation}
where the bound of $W$ is determined  by

\begin{equation}
W ={\rm dim}{\cal H}< \sum^N_{m=0}\frac{z^m}{(m!)^2} \leq
\sum^\infty_{m=0}\frac{z^m}{(m!)^2}=I_0(2\sqrt{z}) \sim
\frac{e^{2\sqrt{z}}}{\sqrt{4\pi\sqrt{z}}}.
\end{equation}

Here $I_0$ is the zeroth-order Bessel function of the second kind.
 Since $z$ is given
by
\begin{equation}
 z= \sum^N_{i=1}L_i \to  \frac{L^3}{2\pi^2}\int_{0}^{\Lambda}\Bigg[\frac{E_{BH}}{p}\Bigg]p^2dp
 =\frac{\Lambda^2L^4}{4\pi^2}.
\end{equation}

Using UV-IR relation of Eq(\ref{UVR}), the bound can be given as follows

\begin{equation}
z  \leq  L^3.
\end{equation}

Since the boundary area of the system is given by

\be
A\sim L^2 \label{area},
\ee
therefore, the bound for the maximum entropy  of Eq(\ref{maxentropy}) will be given by
\begin{equation}
S_{\rm max}=\ln W \leq A^{3/4}.
\end{equation}
This is just a brief summary of determining entropy bound by using the Local Quantum Field Theory (LQFT).

\subsection{The effect of GUP on The Holographic Entropy Bound and LQFT }

Consider a massless bosonic field confined to cubic box of size L ,
as has been done in subsection $3.1$, but now with including the GUP modification.
Using Eq(\ref{Densitystates}), the total number of the quantized modes will be modified as follows,

\begin{equation}
 N \to \frac{L^3}{2\pi^2}\int_{0}^{\Lambda}\frac{p^2dp}{(1-\alpha p)^4}
 \approx
 \frac{L^3}{2\pi^2}\left(\frac{\Lambda^3}{3}+\alpha \Lambda^4\right).
\end{equation}

We note the  total number of states is increased  due to GUP correction.
Note that, this result is valid subject to:
\be
\frac{1}{ \alpha} > \Lambda,
\ee

otherwise, the number of states will be infinite or negative number.
This means $\alpha$ gives a boundary on the cutoff $\Lambda$.\\

Now turning to the modifications implied by GUP on the energy bound up to the first order of $\a$ , we find

\begin{equation}
 E \to \frac{L^3}{2\pi^2}\int_{0}^{\Lambda}\frac{p^3dp}{(1-\alpha p)^4}
 \approx
 \frac{L^3}{2\pi^2}\left(\frac{\Lambda^4}{4}+\alpha\frac{4\Lambda^5}{5}\right)\\
 \leq E_{BH}\label{eneregy2}.
\end{equation}

Using equations (\ref{energy},\ref{UVR}) with the  last inequality (\ref{eneregy2}), we get the following UV-IR relation
up to the first order of $\a$

\begin{eqnarray} &&\frac{L^3}{8\pi^2}\left(\Lambda^4+\alpha\frac{16\Lambda^5}{5}\right)\nonumber
 \leq L,\\  &&\Lambda^4\left(1+\a \frac{16 \Lambda}{5}\right)\leq\frac{1}{L^2}, \nn\\
&&\Lambda^2\leq \frac{1}{L}\left(1-\frac{8\alpha}{5L^{\frac{1}{2}}}\right)\label{Scale}.
\end{eqnarray}

On the other side, the modified maximum entropy has been calculated according
to the following procedure

\begin{equation}
S_{\rm max}=\ln W,
\end{equation}

with  $W\sim e^{2\sqrt{z}}$. Since $z$ is given  up to the first order of $\alpha$ by

\begin{equation}
 z \to  \frac{L^3}{2\pi^2}\int_{0}^{\Lambda}\Bigg[\frac{E_{BH}}{p}\Bigg]\frac{p^2 dp}{(1-\alpha p)^4}
 \approx
 \frac{L^4}{2\pi^2}\left(\frac{\Lambda^2}{2}+\alpha\frac{4\Lambda^3}{3}\right),
\end{equation}
one finds the bound when using  UV-IR relation in Eq.(\ref{Scale})

\bea
&&z \leq \frac{L^4}{4 \pi^2} \left(\Lambda^2+ \a \frac{8 \Lambda^3}{3}\right),\nn\\
&&z \leq \frac{L^4}{4 \pi^2} \left( \frac{1}{L}\big(1-\frac{8\alpha}{5L^{\frac{1}{2}}}\big)+ \a \frac{8}{3 L^{\frac{3}{2}}}\right),
\nn\\
&&z \leq  L^3 +\frac{16\alpha L^{\frac{5}{2}}}{15}.
\eea

Using the boundary area of the system of Eq(\ref{area}), we find the bound for the maximum entropy will be modified as follows,

\begin{equation}
S_{\rm max}=\ln W \leq A^{3/4}+\frac{16 \alpha}{30}A^{1/2}, \label{result}
\end{equation}

which shows clearly that the upper bound is increased due to the
GUP. This means that the the maximum entropy that can be stored in a bounded
region of space has been increased due to the presence of the GUP or, in other words, by
considering the minimal length in Quantum Gravity. This shows that the
conjectured entropy of the truncated Fock space corrected by the GUP disagrees
with 't Hooft's classical result which requires disagreement between the
micro-canonical and canonical ensembles for a system with a large number
of degrees of freedom due to the GUP-correction term. Then the holographic
theory doesn't retain its good features. On the other side, since the GUP implies discreteness
of space by itself as proposed in \cite{Ali:2010yn,advplb,Das:2010zf}, therefore the discreteness of
space will not leave the continuous symmetries such as rotation and Lorentz symmetry intact,
which means by other words the holographic theory doesn't retain
its good features\cite{Susskind:1994vu}. Possibilities of violating Holographic
theory near the Planck scale have been discussed by many authors see e.g \cite{holography}.
This seems that holographic theory does not retain its good features
by considering minimal length in Quantum Gravity.
\newpage

\section{Conclusions}
In this paper we tackle the problem of studying the possible discrepancy that
has been found between the results of neutron interferometry experiment
and Quantum Mechanics. We investigated whether the GUP can explain the violation
of weak equivalence principle at small length scales. We have shown that, by studying
Heisenberg equations of motions in the presence of GUP, the acceleration is no
longer mass-independent because of the mass-dependence through the momentum p.
Therefore, the equivalence principle is dynamically violated.

We also investigated the consequences of the GUP on the Liouville theorem.
We found a new form of an invariant phase space in the presence of the GUP.
In the future, it would be nice to apply our approach on the calculations
of the cosmological constant, black body radiation, etc. We applied our approach
on the calculation of the entropy bound of local quantum field theory.
This led to a $\sqrt{A}$-type correction  to  the bound of the maximal
entropy of a bosonic field. This showed that the conjectured entropy
of the truncated Fock space corrected by GUP disagrees with 't Hooft's
classical result. This agreed with the discreteness of space
implications which  does not leave the continuous symmetries such as translation,
rotation and full Lorentz symmetry intact, and hence the holographic theory
doesn't retain its good features due to discreteness of space.\\

{\large{\bf Acknowledgments} ; }\\ The author gratefully thanks theoretical
physics group at University of Lethbridge for many enlightening
discussions at journal clubs on the subject. The author thank the anonymous referee for useful comments and suggestions. This work was supported in part by the Natural Sciences and Engineering Research Council of Canada
and by  University of Lethbridge.



\end{document}